\documentclass{article}
\usepackage{naturetex}
\usepackage{amsmath}

\begin{document}

\maketitle

{\setstretch{1.0}
	\section*{Abstract}
The duration of the accumulation rate (physical image) is a key factor in analysis of counterintuitive phenomena involving relative timescales on multiplex networks. Typically, the relative timescales are represented by multiplying any layer by the same factor. However, researchers often overlook the changes in the relative timescales caused by local parameters, resulting in incomplete analysis of phenomena. This paper examines the survival time of stifler individuals in the information-epidemic model on multiplex networks. The relative timescales can be affected by the survival time (only one parameter), reversing the monotonically increasing phenomenon into a monotonically decreasing one, that is, a counterintuitive phenomenon under incomplete analysis. Additionally, the relative timescales can influence the epidemic threshold, which is different from the previous studies. Our work suggests that considering the physical image of relative timescales is crucial when analyzing multiplex networks, even when only one parameter is altered.
}

\newpage
\section*{Introduction}
Spreading phenomena exist widely in natural and social systems~\cite{PhysRevLett.130.247401}, and various corresponding models have been established, such as the susceptible-infected-susceptible (SIS) model~\cite{RevModPhys.87.925}, the bass diffusion model~\cite{Zhang2020}, the majority-vote model~\cite{PhysRevE.71.016123}, susceptible-infected-recovered model~\cite{Aguilar2023}, and the threshold model~\cite{PhysRevE.75.036109}.
An information diffusion model, the unaware-aware-unaware model, which is an analogue of the disease spread model, is often  used to study the information-epidemic coupling model on multiplex networks~\cite{PhysRevLett.111.128701,Zhou_2019,PhysRevE.90.012808,KAN2017193,PhysRevResearch.3.013157,XIA2019185,chen2023epidemic,PhysRevResearch.5.033065}. 
However, information diffusion, due to human factors, has two characteristics that differ from traditional disease spread: group interaction (also called higher-order interactions) and individuals who know but do not transmit (called stifler individuals).
The principle of the latter is that some aware individuals find that themselves surrounded by other aware individuals, or that too much time has passed, and then they feel that it is unnecessary to continue spreading the information (thinking most individuals are aware). 
On a single layer network, there are functional similarities between the stifler individuals of the information diffusion model and the removed ones of the epidmeic model~\cite{borge2013emergence} (the way they interact may be different~\cite{daley1964epidemics,Junior_2021,ferraz2022subcritical}), but because the stifler individuals still carry information, it may play an important role for the information-epidemic model on multiplex networks.

Higher-order interactions refer to multiple simultaneous interactions within the same group that produce additional persuasive (or infecting) effects~\cite{aad9029,BATTISTON20201,
battiston2021physics,rosas2022disentangling,BOCCALETTI20231,BOCCALETTI20231,PhysRevResearch.2.023032,PhysRevResearch.5.013201,Santoro2023}. 
For instance, people are more likely to be persuaded by the simultaneous persuasion of two friends than by their solo persuasion.
The researchers studied the higher-order interaction on single-layer networks~\cite{PhysRevResearch.2.012049,iacopini2019simplicial,Chowdhary_2021,doi:10.1063/5.0040518,PhysRevLett.127.158301,PhysRevE.104.044303} and also introduced it into the information-epidemic model on multiplex networks~\cite{10.1063/5.0099183,doi:10.1098/rspa.2022.0059,10.1063/5.0125873,PhysRevResearch.5.013196,LIU2023113657}.
Introducing higher-order interactions can effectively inhibit disease spread and increase the epidemic threshold~\cite{doi:10.1098/rspa.2022.0059,LIU2023113657}.
The combination of simplicial complexes and self-awareness has a significant impact on the epidemic layer, for example, the threshold must be more carefully identified from the multiple peaks of the susceptibility curve~\cite{PhysRevResearch.5.013196}.

Recently, the study of the relative timescales between different layers on multiplex networks has become increasingly attractive due to the emergence of many counterintuitive results~\cite{Wang_2017_njp,PhysRevE.100.032313,PhysRevE.102.022312,PhysRevResearch.5.033220,PhysRevResearch.3.013146}.
Wang $et\ al.$ found that faster information diffusion does not always better mitigate the epidemic spreading~\cite{Wang_2017_njp}.
Ventura da Silva $et\ al.$ indicate that the epidemic prevalence increases with the speed of information processing on the information-epidemic model on multiplex networks~\cite{PhysRevE.100.032313,PhysRevE.102.022312}.
Cai $et\ al.$ revealed the mysteries of relative timescales in cyclic and irreversible dynamics using physical images.
One of these images, the discretization of the low rate region, shows that the average time required to accumulate unit infection rate decreases as the relative timescales increase~\cite{PhysRevResearch.5.033220}.

However, there are still potential worries regarding relative timescales on multiplex networks.
Previous studies have required that all parameters within layers be changed simultaneously~\cite{PhysRevE.100.032313,PhysRevE.102.022312,PhysRevResearch.5.033220}, but we have found that a single parameter can also affect the relative timescales between the layers.
While Refs.~\cite{PhysRevE.100.032313,PhysRevE.102.022312} theoretically shows that the epidemic threshold seems not to be affected by the relative timescales, we demonstrate that the epidemic threshold is still influenced by them.
The above understanding is found in this paper by analyzing the impact of stifler individuals on the information-epidemic model, in which the reduced survival time of stifler individuals is a ``double-edged sword".

\newpage
\section*{Results}\label{results}
\subsection*{Information epidemic model with stifler individuals and higher-order interactions}

Multiplex networks are utilized to investigate the coupling information-epidemic dynamics. 
The upper layer represents the information layer and employs the simplicial unaware-aware-stifler-unaware (sUALU) model~\cite{doi:10.1063/5.0040518}, while the lower layer represents the epidemic layer and uses the SIS model.
The sUALU-SIS model consists of six possible states for individuals: unaware and susceptible (US), aware and susceptible (AS), stifler and susceptible (LS), unaware and infected (UI), aware and infected (AI), and stifler and infected (LI).

In the information layer, there are three possible state transitions, from unaware to aware (U$\to$A), from aware to stifler (A$\to$L), and from stifler to unaware (L$\to$U), which form a loop.
For the U$\to$A process, the epidemic-related information is diffused at the rate of $\alpha$ by links $[n_{1}(UY),n_{2}(AY)]$ (where $Y$ is either $S$ or $I$) and at the rate of $\upsilon_{\Delta}\alpha$ by $2-$simplices $[n_{1}(UY),n_{2}(AY),n_{3}(AY)]$, where $2-$simplex is a way of characterizing group interaction (higher-order interaction). 
The A$\to$L process involves an aware individual becoming a stifler who loses interest in propagating the information at a rate of $\delta$.
The L$\to$U process involves a stifler individual forgetting the information and returning to being unaware at the rate of $\gamma$.
The possible events of the information layer are as follows,
\begin{eqnarray}
	\left[n_{1}(UY),n_{2}(AY)\right]&\stackrel{\alpha}{\longrightarrow}&\left[n_{1}(AY),n_{2}(AY)\right], \nonumber \\	
	\left[n_{1}(UY),n_{2}(AY),n_{3}(AY)\right]&\stackrel{\upsilon_{\Delta}\alpha}{\longrightarrow}&\left[n_{1}(AY),n_{2}(AY),n_{3}(AY)\right], \nonumber \\
	\left[n_{1}(AY)\right]&\stackrel{\delta}{\longrightarrow}&\left[n_{1}(LY)\right], \nonumber \\
	\left[n_{1}(LY)\right]&\stackrel{\gamma}{\longrightarrow}&\left[n_{1}(UY)\right]. \nonumber 
\end{eqnarray}

In the epidemic layer, a susceptible individual who does not know the information is infected by contacting with an infected individual at a rate of $\beta$, while the infection rate changes to $\eta\beta$ as a result of protective measures taken by the susceptible individual when the information is known.
Here, $\eta\in[0,1]$ relates to the strength of protective measures taken by aware individuals.
Infected individuals recover to be susceptible status at a rate of $\mu$.
The potential events in the epidemic layer are as follows,
\begin{eqnarray}
	\left[n_{1}(US),n_{2}(XI)\right]&\stackrel{\beta}{\longrightarrow}&\left[n_{1}(UI),n_{2}(XI)\right], \nonumber \\	
	\left[n_{1}(AS),n_{2}(XI)\right]&\stackrel{\eta\beta}{\longrightarrow}&\left[n_{1}(AI),n_{2}(XI)\right], \nonumber \\
	\left[n_{1}(XI)\right]&\stackrel{\mu}{\longrightarrow}&\left[n_{1}(XS)\right]. \nonumber 
\end{eqnarray}
where $X\in\{U,A,L\}$. 

In multiplex networks, the neighbors for any individual maybe different in the two layers.
We employ the random simplicial complex network~\cite{iacopini2019simplicial,PhysRevResearch.5.013196} and the ER random network in the information layer and the epidemic layer, respectively.
Our model does not consider self-awareness~\cite{PhysRevResearch.5.013196}, but it is suitable for some infectious diseases with no obvious or mild symptoms. For example, most people infected with Toxoplasma gondii are asymptomatic carriers~\cite{doi:10.3109/00365548.2012.693197}.
Ignoring self-awareness can make the contrast clearer. In particular, changing the parameters of the epidemic layer does not directly affect the information layer but changes the relative timescales between the two layers, while changing the parameters of the information layer not only affects the epidemic layer through the interlayer interaction but also changes the relative timescales.

This paper focuses on the impact of higher-order interactions $\upsilon_{\Delta}$, forgetting rate $\gamma$ and recovery rate $\mu$ (or infection rate $\beta$) on the coupled information-epidemic dynamics.
Specifically, we examine how a single parameter can affect the relative timescales between the layers.
In this paper, we fix population size $N=10000$, the average degree of the information layer $\langle k_1\rangle=20$, the expected number of 2-simplices $\langle k_{\Delta}\rangle=6$, and the average degree of the epidemic layer $\langle k_2\rangle=6$. 
Meanwhile, we fix $\eta=0$ to maximize the impact of the information on the epidemic layer. 
To prevent bistable in prevalence and periodic oscillations~\cite{doi:10.1063/5.0040518}, we use $\delta=0.2$, $\alpha=0.02$, and $\gamma\in[0.05,1]$.
The choice of parameters makes the analysis more intuitive.

\subsection*{Far from the epidemic threshold}

\begin{figure}
	\includegraphics[width=\textwidth]{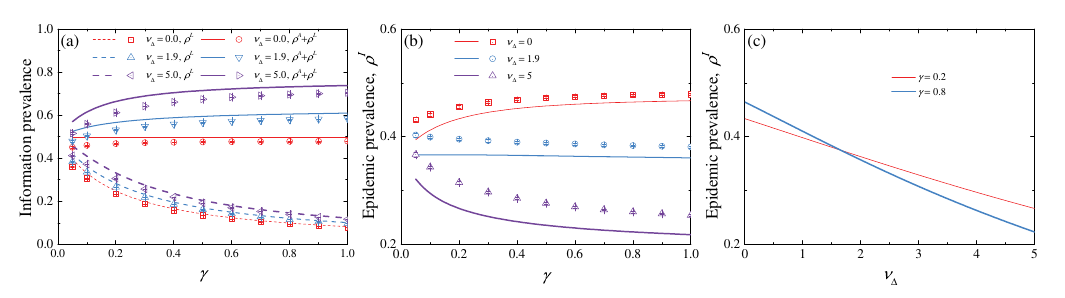}
	\caption{(Color online)  {\bf The effect of the average survival time of stifler individuals ($1/\gamma$) on the information-epidemic model.}
(a) The information prevalence $\rho^{L}$ and $(\rho^{A}+\rho^{L})$ as a function of $\gamma$ for different $\upsilon_{\Delta}$. (b) The epidemic prevalence $\rho^{I}$ as a function of $\gamma$ for different $\upsilon_{\Delta}$. (c) The epidemic prevalence $\rho^{I}$ as a function of $\upsilon_{\Delta}$ for different $\gamma$. The scatters are simulation results obtained by averages from $100$ independent runs, with error bars indicating the standard deviation from the ensemble  mean. The theoretical results, shown as solid and dashed lines, can be obtained from Eqs.~\ref{t4}-\ref{t7}. In panel (a), the theoretical results can also be obtained more easily from Eq.~\ref{t3} or \ref{t33}. Parameters: $N=10000$, $\eta=0$, $\delta=0.2$, $\alpha=0.02$, $\beta=0.8$, $\mu=0.5$, $\rho^{A}(0)=0.05$, $\rho^{I}(0)=0.05$ and $\rho^{L}(0)=0$.\label{f3}}
\end{figure}

The increase in $\gamma$ means a shorter survival time for stifler individuals, leading to a decrease in $\rho^{L}$ as shown in Fig.~\ref{f3} (a).
Interestingly, however, this process aids in the transmission of information, as shown by the increase in $\rho^{A}+\rho^{L}$ with increasing $\gamma$, as seen in Fig.~\ref{f3} (a).
When considering the mean field approximation and $\upsilon_{\Delta}=0$ (represented by the thin red line in Fig.~\ref{f3} (a)), Eq.~\ref{t33} shows that $\rho^{A}+\rho^{L}$ is independent of $\gamma$. This indicates that the decrease in $\rho^{L}$ and the increase in $\rho^{A}$ are precisely equal as $\gamma$ increases.
The red circle simulation results in Fig.~\ref{f3} (a) show a slight increase as $\gamma$ increases, which is due to the tendency of aware individuals to group together in static networks (called dynamical correlation in Refs.~\cite{PhysRevLett.116.258301,PhysRevE.103.032313}). Specifically, when a stifler individual becomes unaware, the proportion of aware individuals in its neighbors is slightly higher than average.
In this paper, the average degree is high enough to make the impact of dynamic correlation relatively insignificant.
When $\upsilon_{\Delta}$ is greater than zero, the increase in $\rho^{A}$ due to additional higher-order interactions is greater than the decrease in $\rho^{L}$, resulting in an increase in $\rho^{A}+\rho^{L}$ as $\gamma$ increases.

\begin{figure}
\centering
	\includegraphics[width=0.5\textwidth]{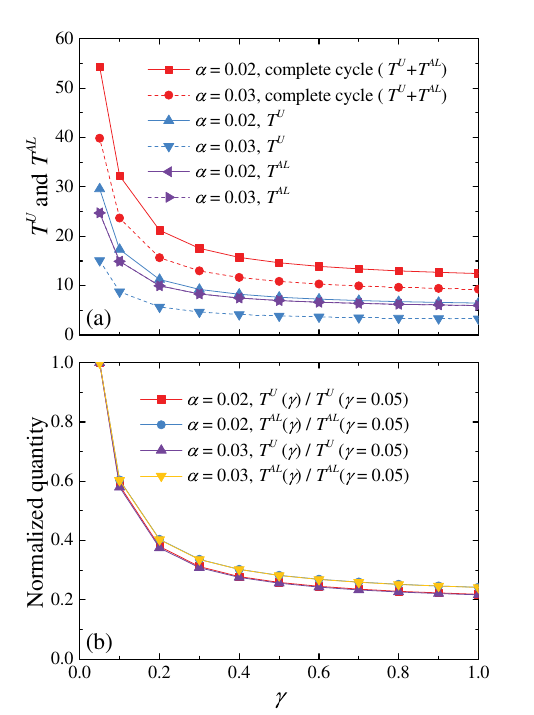}
	\caption{(Color online) {\bf Scaling of the diffusion speed of the information layer} (a) The average duration of state as a function of $\gamma$. (b) Values of $T^U$ and $T^{AL}$, normalized by their values at the $\gamma=0.05$, as a function of $\gamma$. The scatters are simulation results obtained by averages from $100$ independent runs. Parameters: $N=10000$, $\eta=0$, $\delta=0.2$, $\beta=0.8$, $\upsilon_{\Delta}=0$, $\mu=0.5$, $\rho^{A}(0)=0.05$, $\rho^{I}(0)=0.05$ and $\rho^{L}(0)=0$.\label{f4}}
\end{figure}

In Fig.~\ref{f3} (b), as $\gamma$ increases, $\rho^{I}$ shows distinct behaviors for various $\upsilon_{\Delta}$, such as either a monotonic increase or decrease.
This phenomenon may seem counterintuitive considering the monotonicity of $\rho^{A}+\rho^{L}$ in Fig.~\ref{f3} (a), but it can be explained by physical images of the relative timescales~\cite{PhysRevResearch.5.033220}.
In order to better show the competition between the two physical mechanisms above, we plot $\rho^{I}$ as a function of $\upsilon_{\Delta}$ for different $\gamma$ in Fig.~\ref{f3} (c).
A clear crossover can be observed in Fig.~\ref{f3} (c): an increase in $\gamma$ promotes disease spread for low $\upsilon_{\Delta}$, while it hinders disease spread for high $\upsilon_{\Delta}$.

\begin{figure}
\centering
	\includegraphics[width=0.5\textwidth]{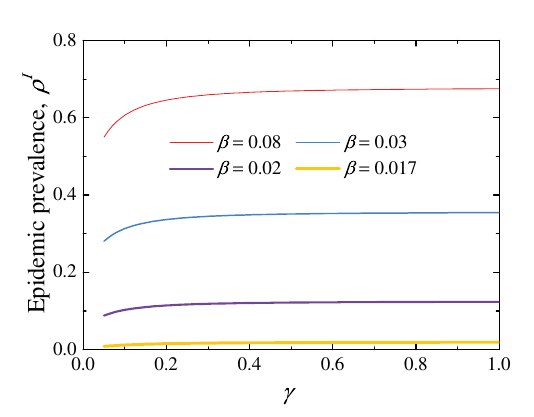}
	\caption{(Color online) {\bf The effect of the relative time scale is influenced by the density of infected individuals.} The epidemic prevalence $\rho^{I}$ as a function of $\gamma$ for different $\beta$. The lines are the theoretical results obtained from Eqs.~\ref{t4}-\ref{t7}. Parameters: $N=10000$, $\eta=0$, $\delta=0.2$, $\alpha=0.02$, $\rho^{A}(0)=0.05$, $\rho^{I}(0)=0.05$ and $\rho^{L}(0)=0$, $\upsilon_{\Delta}=0$, $\mu=0.05$.
\label{f2}}
\end{figure}

To intuitively illustrate the effect of $\gamma$ on the speed of information processing, as shown in Fig.~\ref{f4} (a), we calculate the average duration of the unaware state ($T^U$) and the sum of the average duration of the aware and stifler states ($T^{AL}$) after the system reached its steady state.
The red scatters in Fig.~\ref{f4} (a) clearly shows that as $\gamma$ increases, the average duration required for a complete cycle ($T^U+T^{AL}$) decreases, indicating an increase in the speed of information processing.
The physical mechanism is easily understandable: the shortening of the average duration of stifler state (i.e., $1/\gamma$) is accompanied by the increasing of aware individuals [see the discussion in Fig.~\ref{f3} (a)], which makes the average duration of unaware state shorter.
In Fig.~\ref{f4} (b), we plot the relative timescales as a function of $\gamma$. 
It is clear that $T^U$ and $T^{AL}$ are discretized in nearly equal proportions.
Based on the physical image of discretization of the low rate region in Ref.~\cite{PhysRevResearch.5.033220}, $\rho^{I}$ monotonically increases for $\upsilon_{\Delta}=0$ in Fig.~\ref{f3} (b) due to the decrease in the average time required to accumulate unit infection rates, which promotes the spread of the disease.
However, for the high $\upsilon_{\Delta}$ in Fig.~\ref{f3} (b), the monotonic decrease in $\rho^{I}$ is due to the increase in $\rho^{A}+\rho^{L}$, resulting in a decrease in the average infection rate generated per unit time.

Finally, Fig.~\ref{f2} demonstrates that the impact of the relative timescales decreases as $\beta$ decreases.
An individual that can be affected by relative timescales must meet two conditions: on the one hand, it must be able to switch between known and unknown information; on the other hand, it must have at least one infected neighbor.
Therefore, the effect of relative timescales decreases as the density of infected individuals decreases. 
Meanwhile, the decrease of gap between $\beta$ and $\eta\beta$ reduces the difference between the high- and the low-rate regions, thus diminishing the effect of relative timescales.

\subsection*{Epidemic threshold}

The epidemic threshold is the point at which the epidemic prevalence transitions from zero to non-zero during steady state.
To determine the epidemic threshold in Monte Carlo simulations, we utilize two commonly used methods: the lifetime-based method (LT)~\cite{PhysRevLett.111.068701} and the quasi-stationary simulation method (QS)~\cite{PhysRevE.86.041125,PhysRevE.94.042308}.
The former is based on the long relaxation time of the order parameter, while the latter is based on the large fluctuations.

\begin{figure}
\centering
	\includegraphics[width=0.5\textwidth]{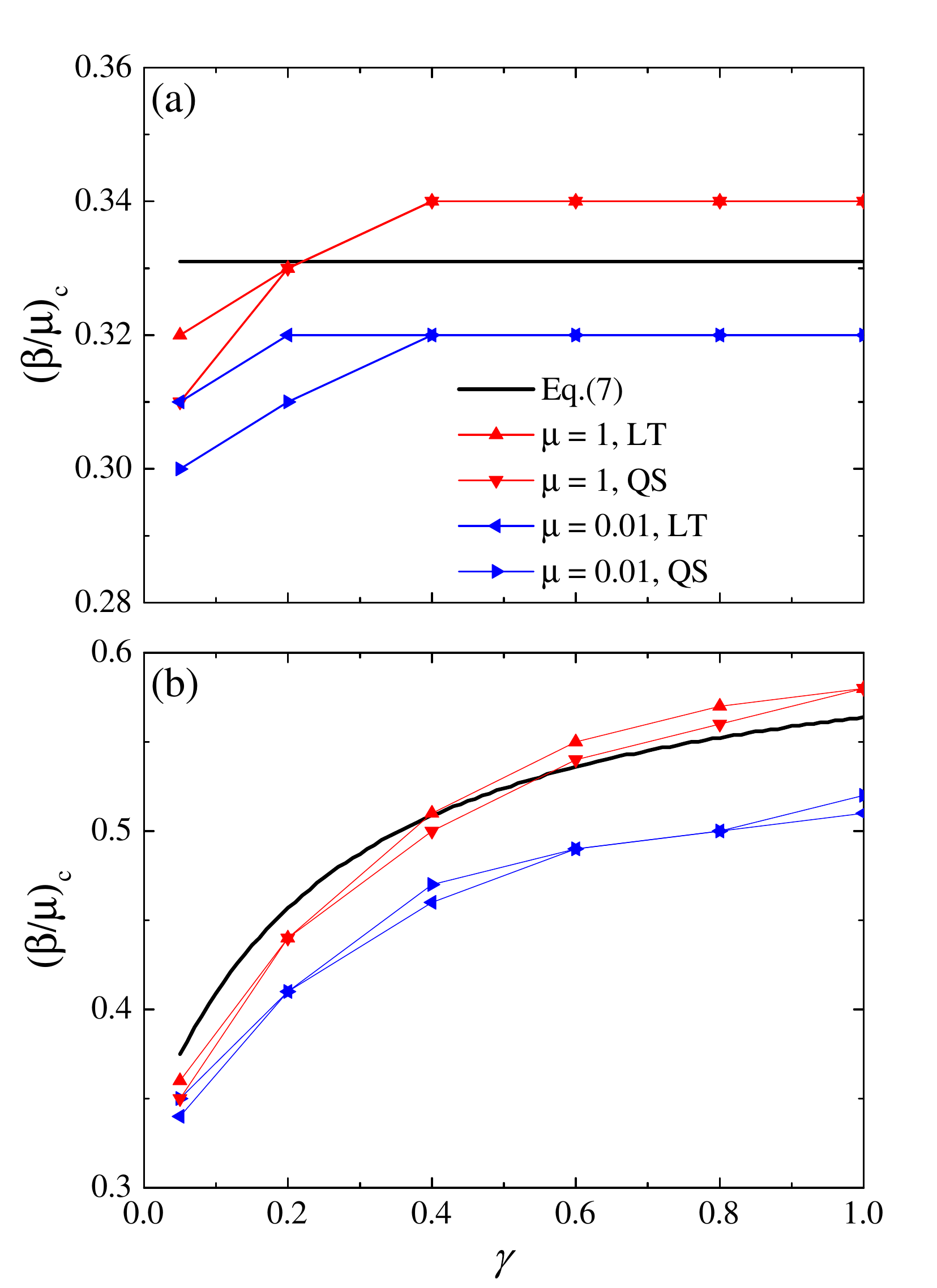}
	\caption{(Color online) {\bf The epidemic thresholds are influenced by relative timescales.} The epidemic threshold as a function of $\gamma$ for different $\upsilon_{\Delta}$ and $\mu$. The lines are the theoretical results obtained from Eq.~\ref{t8}, while scatters are simulation results. Parameters: $N=10000$, $\eta=0$, $\delta=0.2$, $\alpha=0.02$; (a) $\upsilon_{\Delta}=0$; (b) $\upsilon_{\Delta}=4$.
\label{f5}}
\end{figure}

\begin{figure}
\centering
	\includegraphics[width=0.5\textwidth]{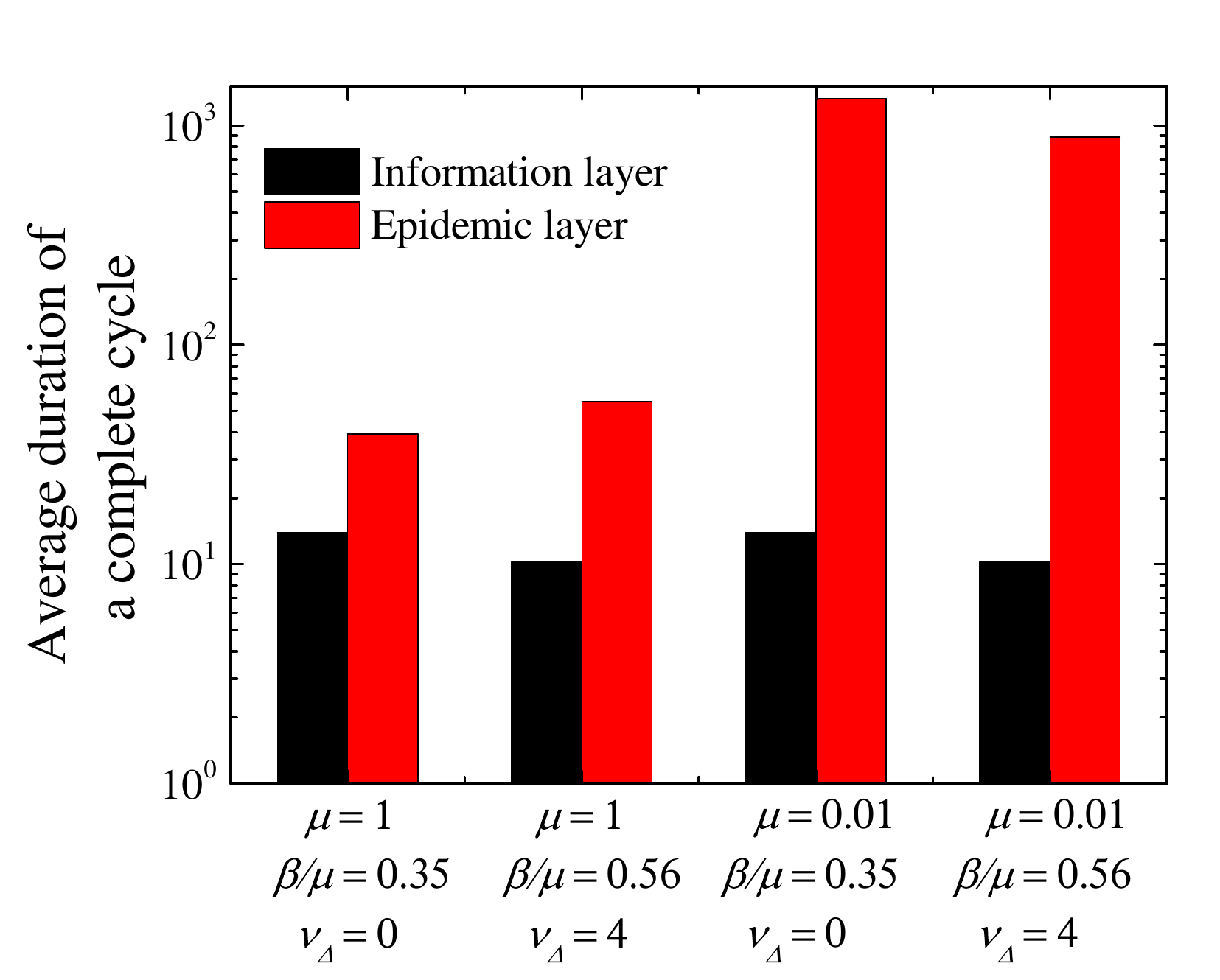}
	\caption{(Color online) {\bf The reduction in recovery rates slowed the spread of the disease.} The average duration of the complete cycle both in the information layer and in the epidemic layer. Parameters: $N=10000$, $\eta=0$, $\delta=0.2$, $\alpha=0.02$, $\gamma=0.6$.
\label{f6}}
\end{figure}

In Fig.~\ref{f5}, we plot the epidemic threshold $(\beta/\mu)_c$ as a function of $\gamma$ for different $\upsilon_{\Delta}$ and $\mu$. 
It is clear that the monotonicity of the threshold in Fig.~\ref{f5} is the same as that of $\rho^{A}+\rho^{L}$ in Fig.~\ref{f3}(a), which has been captured by Eq.~\ref{t8a}.
It is important to consider whether the relative timescales affect the epidemic threshold, but Fig.~\ref{f4}(b) shows that the relative timescales reach saturation as $\gamma$ increases.
Thus, in Fig.~\ref{f5}, we increase the relative timescales by decreasing $\mu$, which slows the spread of the disease.
Figure~\ref{f6} displays the average duration of the complete cycle in both the information and epidemic layers. 
As shown in Fig.~\ref{f6}, decreasing $\mu$ increases the average duration of the disease layer increases by two orders of magnitude, significantly reducing the speed of the disease spread.
By combining Fig.~\ref{f5} and Fig.~\ref{f6}, it is observed that an increase in relative timescales leads to a lower threshold.

The theoretical results in Fig.~\ref{f5} slightly differ from the simulation results for two reasons: Eq.~\ref{t8a} only considers the average infection rate, which ignores the effect of relative timescales; Eq.~\ref{t8b} ignores the dynamic correlation of the information layer. 

\newpage
\section*{Discussion}

In summary, we have investigated the effect of the average survival time ($1/\gamma$) of stifler individuals, who possess information but have lost interest in diffusing it, on the sUALU-SIS epidemic-information model.
We are surprised to find that changing just one parameter, the forgetting rate $\gamma$, significantly altered the relative timescales of layers.

For far from the epidemic threshold, we found some interesting behaviors.
The survival time of stifler individuals is shortened, which directly reduces the fraction of stifler individuals $\rho^{L}$ but indirectly leads to an increase in the fraction of aware individuals $\rho^{A}$, resulting in an increase of $\rho^{A}+\rho^{L}$ and ultimately promoting the transmission of information.
Meanwhile, for the epidemic layer, shortening the survival time of stifler individuals can be a ``double-edged sword", as it may either increase or decrease the fraction of infected individuals $\rho^{I}$.
This effect is caused by two competing physical mechanisms. 
On the one hand, the increase in $\rho^{A}+\rho^{L}$ results in a decrease in the average infection rate per unit time.
On the other hand, increasing $\gamma$ accelerates information processing speed, reducing the average time required for the accumulation of unit infection rate, as shown by the physical images of the relative timescales.
Additionally, we find that the physical mechanism associated with the relative timescales weakens as the fraction of infected individuals decreases.

For the epidemic threshold, the monotonicity of the threshold is the same as that of $\rho^{A}+\rho^{L}$, regardless of whether the group interaction is considered or not.
Theoretical and simulation results support this conclusion.
Meanwhile, we confirm that when the recovery rate is decreased by two orders of magnitude, changing the relative timescales, the threshold is lowered.

Our dynamic equations serve a baseline only, allowing for more intuitive analysis of the results. 
If the mean-field approach is replaced with the effective degree approach in the information layer~\cite{Zhou_2019}, which considers the dynamic correlation and degree distribution in the information layer, the theoretical results will be more consistent with the simulation results.
Currently, there is still lack of the threshold formula that can account for changes in relative timescales, and it is a challenge for incorporating the average time required for the accumulation of unit infection rates into the theoretical analysis.
As a final remark, we anticipate that our study will open the door for complete analysis  of other phenomena in dynamic models of multiplex networks, both in terms of average rate per unit time and cumulative unit rate duration.

\newpage
\section*{Methods}
\subsection*{The dynamic equations of the sUALU-SIS model on multiplex networks}
Here, the mean-field approach from Ref.~\cite{doi:10.1063/5.0040518} is adopted for the information layer, and the effective degree approach from Ref.~\cite{lindquist2011effective} is adopted for the epidemic layer.
Individuals are classified as $XY_{s,i}$, where $X\in\{U,A,L\}$ and $Y\in\{S,I\}$. The subscripts $s$ and $i$ represent the numbers of susceptible neighbors and infected neighbors in the epidemic layer, respectively.
The dynamic equations of the sUALU-SIS model are as follows:
\begin{subequations}\label{t4}
\begin{eqnarray}\label{t41}
	\frac{dUS_{s,i}}{dt}&=&-US_{s,i}\beta i-US_{s,i}Gs+US_{s+1,i-1}G(s+1)+UI_{s,i}\mu-US_{s,i}i\mu+US_{s-1,i+1}\mu(i+1)+LS_{s,i}\gamma  \nonumber\\
&&-US_{s,i}\alpha\langle k_1\rangle\rho^{A}-US_{s,i}\langle k_{\Delta}\rangle\upsilon_{\Delta}\alpha(\rho^{A})^{2},  \\ \label{t42}
	\frac{dUI_{s,i}}{dt}&=&US_{s,i}\beta i-UI_{s,i}Hs+UI_{s+1,i-1}H(s+1)-UI_{s,i}\mu-UI_{s,i}i\mu+UI_{s-1,i+1}\mu(i+1)+LI_{s,i}\gamma  \nonumber\\
&&-UI_{s,i}\alpha\langle k_1\rangle\rho^{A}-UI_{s,i}\langle k_{\Delta}\rangle\upsilon_{\Delta}\alpha(\rho^{A})^{2}, \\ \label{t43}
	\frac{dAS_{s,i}}{dt}&=&-AS_{s,i}\eta\beta i-AS_{s,i}Gs+AS_{s+1,i-1}G(s+1)+AI_{s,i}\mu-AS_{s,i}i\mu+AS_{s-1,i+1}\mu(i+1)-AS_{s,i}\delta  \nonumber\\
&&+US_{s,i}\alpha\langle k_1\rangle\rho^{A}+US_{s,i}\langle k_{\Delta}\rangle\upsilon_{\Delta}\alpha(\rho^{A})^{2},   \\ \label{t44}
	\frac{dAI_{s,i}}{dt}&=&AS_{s,i}\eta\beta i-AI_{s,i}Hs+AI_{s+1,i-1}H(s+1)-AI_{s,i}\mu-AI_{s,i}i\mu+AI_{s-1,i+1}\mu(i+1)-AI_{s,i}\delta  \nonumber\\
&&+UI_{s,i}\alpha\langle k_1\rangle\rho^{A}+UI_{s,i}\langle k_{\Delta}\rangle\upsilon_{\Delta}\alpha(\rho^{A})^{2},  \\ \label{t45}
	\frac{dLS_{s,i}}{dt}&=&-LS_{s,i}\eta\beta i-LS_{s,i}Gs+LS_{s+1,i-1}G(s+1)+LI_{s,i}\mu-LS_{s,i}i\mu+LS_{s-1,i+1}\mu(i+1)+AS_{s,i}\delta  \nonumber\\ 
&&-LS_{s,i}\gamma,\\ \label{t46}
	\frac{dLI_{s,i}}{dt}&=&LS_{s,i}\eta\beta i-LI_{s,i}Hs+LI_{s+1,i-1}H(s+1)-LI_{s,i}\mu-LI_{s,i}i\mu+LI_{s-1,i+1}\mu(i+1)+AI_{s,i}\delta \nonumber\\
&&-LI_{s,i}\gamma.
\end{eqnarray}
\end{subequations}
The terms $US_{s,i}\beta i$, $AS_{s,i}\eta\beta i$, and $LS_{s,i}\eta\beta i$ represent the rate at which a susceptible individual of the corresponding class is infected by infected neighbors.
The term $XI_{s,i}\mu$ represents the rate that infected individuals recover spontaneously at rate $\mu$. 
The term $XS_{m,n}Gm$ (or $XI_{m,n}Hm$) represents the rate that individuals in $XS_{m,n}$ (or $XI_{m,n}$) leave the class and enter $XS_{m-1,n+1}$ (or $XI_{m-1,n+1}$). Here, $G$ (or $H$) denotes the average infection rate of susceptible neighbors in susceptible (or infected) individuals and can be calculated by
\begin{eqnarray}
	&&G=\frac{\sum_{s,i}(US_{s,i}\beta i+AS_{s,i}\eta\beta i+LS_{s,i}\eta\beta i)s}{\sum_{s,i}(US_{s,i}+AS_{s,i}+LS_{s,i})s},  \nonumber\\
	&&H=\frac{\sum_{s,i}(US_{s,i}\beta i+AS_{s,i}\eta\beta i+LS_{s,i}\eta\beta i)i}{\sum_{s,i}(US_{s,i}+AS_{s,i}+LS_{s,i})i}.  \nonumber
\end{eqnarray}
The term $XY_{m,n}n\mu$ represents the rate that individuals in $XY_{m,n}$ leave the class and enter $XY_{m+1,n-1}$.
The term $LY_{s,i}\gamma$ represents the rate that individuals go from stifler to unaware at rate $\gamma$.
The term $AY_{s,i}\delta$ represents the rate that individuals go from aware to stifler at rate $\delta$.
The terms $UY_{s,i}\alpha\langle k_1\rangle\rho^{A}$ and $UY_{s,i}\langle k_{\Delta}\rangle\upsilon_{\Delta}\alpha(\rho^{A})^{2}$ represent the average rate at which unaware individuals are informed by aware neighbors and group interactions, respectively.

When the population reaches its steady state, the fractions of aware individuals, stifler individuals, and infected individuals in the sUALU-SIS model can be obtained by
\begin{eqnarray}\label{t7}
	\rho^{A}&=&\sum_{s,i}[AS_{s,i}(\infty)+AI_{s,i}(\infty)],  \nonumber\\
	\rho^{L}&=&\sum_{s,i}[LS_{s,i}(\infty)+LI_{s,i}(\infty)],  \nonumber\\
	\rho^{I}&=&\sum_{s,i}[UI_{s,i}(\infty)+AI_{s,i}(\infty)+LI_{s,i}(\infty)].  
\end{eqnarray}

\subsubsection*{The threshold for the sUALU-SIS model}
According to the relationship between the threshold on single-layer networks and that on multiplex networks in Ref.~\cite{PhysRevE.104.044303}, the analysis of the sUALU-SIS model on multiplex networks is as follows.
When the system is in its steady state, there are $N\rho^{U}$ unaware individuals who do not know the information and $N(\rho^{A}+\rho^{L})$ individuals who know the information.
Those susceptibles who are unaware are infected at a rate of $\beta$, while the remaining susceptible are infected at a rate of $\eta\beta$.
Since the information layer is in thermodynamic equilibrium, the average infection rate from a link S-I in the epidemic layer is $\beta[1-(1-\eta)(\rho^{A}+\rho^{L})]$.
Then, for the sUALU-SIS model we have the following relation
\begin{eqnarray}\label{t11}
\left(\frac{\beta}{\mu}\right)_c&=&\frac{1}{1-(1-\eta)(\rho^{A}+\rho^{L})}\left(\frac{\beta}{\mu}\right)_c^{\text{single}},
\end{eqnarray}
where $\left(\frac{\beta}{\mu}\right)_c^{\text{single}}$ represents the epidemic threshold for the single epidemic layer.

In this paper, we use the dynamic correlation method~\cite{PhysRevLett.116.258301} to calculate the threshold on single-layer networks, which considers the dynamic correlation of the immediate neighbors and ignores the higher-order neighbors. The relation is as follows
\begin{eqnarray}\label{t111}
\left(\frac{\beta}{\mu}\right)_c^{\text{single}}=\frac{\langle k_2\rangle}{\langle k_2^{2}\rangle-\langle k_2\rangle}.
\end{eqnarray}

By adding Eq.~\ref{t43} and Eq.~\ref{t44} [Eq.~\ref{t45} and Eq.~\ref{t46}] and then summing the subscripts $s$ and $i$, we get
\begin{eqnarray}\label{t1}
	\frac{d\rho^{A}}{dt}&=&\alpha(1-\rho^{A}-\rho^{L})\langle k_1\rangle\rho^{A}-\delta\rho^{A}+\upsilon_{\Delta}\alpha(1-\rho^{A}-\rho^{L})\langle k_{\Delta}\rangle(\rho^{A})^{2}, \nonumber \\
	\frac{d\rho^{L}}{dt}&=&\delta\rho^{A}-\gamma\rho^{L},  
\end{eqnarray}
which is the same as the mean field approach of the simplicial SIRS model~\cite{doi:10.1063/5.0040518}.
When the population reaches its steady state, the nontrivial solution of Eq.~\ref{t1} is
\begin{eqnarray}\label{t3}
	\rho^{A}&=&\frac{[\gamma\left(\upsilon_{\Delta}\langle k_{\Delta}\rangle-\langle k_1\rangle\right)-\delta\langle k_1\rangle]+\sqrt{\Delta}}{2\upsilon_{\Delta}\langle k_{\Delta}\rangle(\delta+\gamma)}, \nonumber \\
	\rho^{L}&=&\frac{\delta}{\gamma}\rho^{A}, \nonumber \\
	\Delta&=&\alpha^{2}[(\delta+\gamma)\langle k_1\rangle+\gamma\upsilon_{\Delta}\langle k_{\Delta}\rangle]^{2}-4\upsilon_{\Delta}\langle k_{\Delta}\rangle\alpha\gamma\delta(\gamma+\delta).
\end{eqnarray}
Combining Eq.~\ref{t11}, Eq.~\ref{t111}, and Eq.~\ref{t3}, we can give the threshold expression of the theoretical analysis for the sUALU-SIS model on multiplex networks as follows
\begin{subequations}\label{t8}
\begin{eqnarray}\label{t8a}
\left(\frac{\beta}{\mu}\right)_c&=&\frac{\langle k_2\rangle}{\langle k_2^{2}\rangle-\langle k_2\rangle}\frac{1}{1-(1-\eta)(\rho^{A}+\rho^{L})},\\ \label{t8b}
\rho^{A}+\rho^{L}&=&\frac{\upsilon_{\Delta}\langle k_{\Delta}\rangle-\langle k_1\rangle}{2\upsilon_{\Delta}\langle k_{\Delta}\rangle}+\frac{\sqrt{\Delta}-\alpha\langle k_1\rangle\delta}{2\upsilon_{\Delta}\alpha\langle k_{\Delta}\rangle\gamma},\\ \label{t8c}
\Delta&=&\alpha^{2}[(\delta+\gamma)\langle k_1\rangle+\gamma\upsilon_{\Delta}\langle k_{\Delta}\rangle]^{2}-4\upsilon_{\Delta}\langle k_{\Delta}\rangle\alpha\gamma\delta(\gamma+\delta). 
\end{eqnarray}
\end{subequations}
Here, there is a divergence point in Eq.~\ref{t8} when $\upsilon_{\Delta}=0$. For the case of $\upsilon_{\Delta}\to0$, we have $\sqrt{\Delta}= x^{\frac{1}{2}}-\frac{1}{2}yx^{-\frac{1}{2}}$, where  $x=\alpha^{2}[(\delta+\gamma)\langle k_1\rangle+\gamma\upsilon_{\Delta}\langle k_{\Delta}\rangle]^{2}$, $y=4\upsilon_{\Delta}\langle k_{\Delta}\rangle\alpha\gamma\delta(\gamma+\delta)$, and the higher-order terms of $\frac{y}{x}$ are ignored.
And then, $(\rho^{A}+\rho^{L})$ can be simplified as 
\begin{eqnarray}\label{t33}
\rho^{A}+\rho^{L}=1-\frac{\delta}{\alpha\langle k_1\rangle}, 
\end{eqnarray}
which can also be obtained as a nontrivial solution by inserting $\upsilon_{\Delta}=0$ into Eq.~\ref{t1}.

\subsubsection*{Simulation procedures of the epidemic threshold}
The simulation details of the lifetime-based method are as follows.
The lifetime $L$ is defined as the time elapsed before the disease either disappears or spreads to a finite fraction $C$ of the network, where $C$ represents the fraction of distinct nodes ever infected during the simulation.
Given the recovery rate $\mu$, we measure the average lifetime of finite realizations $\bar{L}(\beta/\mu)$, and the position of its peak corresponding to $\beta/\mu$ is an estimate of the threshold for the finite population.
This paper obtains each threshold point by conducting $100$ independent realizations and setting $C$ to $0.9$.
In the information-disease model, prevalence may peak before reaching a steady state~\cite{PhysRevResearch.5.013196} due to the gradual increase of the inhibitory effect from the information layer.
Therefore, to accurately capture thresholds, we must ensure that the information layer reach to a stable state before the epidemic strain infects an initial number of individuals.

The details of the quasi-stationary simulation method are as follows.
We use the reflecting boundary condition which has equivalent threshold predictions to the standard quasistationary simulation method~\cite{PhysRevE.94.042308}. The reflecting boundary means that when any layer enters the absorbing state, the dynamics return to its immediately previous configuration. 
After a long relaxation time $t_{r}$, we capture samples of the number of infected individuals during a period of time $t_{a}$. 
The probability $P(n)$ that the system has $n$ infected individuals can be determined through these samples.
The moments of the activity distribution can be expressed as $\langle(\rho^I)^k\rangle=\sum_{n}(n/N)^{k}P(n)$.
The susceptibilities $\chi^I$ can then be obtained by $\chi^I=N[\langle(\rho^I)^{2}\rangle-\langle\rho^I\rangle^{2}]/\langle\rho^I\rangle$, and the position of the peak, corresponding to $\beta/\mu$, provides an estimate of the threshold for the finite population.
In this paper, we set $t_{r}=2\times10^5$ and $t_{a}=4\times10^5$.

\subsection*{Code availability}
All relevant computer codes are available are available from the corresponding author upon reasonable request.

\subsection*{Data availability}
The data that support the findings of this study are available from the corresponding author upon reasonable request.

\newpage
\bibliography{bibliography}
\bibliographystyle{naturemag}

\newpage
\section*{End Notes}
\subsection*{Acknowledgements}
This work was supported by the Shaanxi Fundamental Science Research Project for Mathematics and Physics (Grant Nos. 22JSQ003 and 22JSZ005) and the National Natural Science Foundation of China (Grant No. 12247103).

\subsection*{Author Contributions}
X.C., C.-R.C., and W.-L.Y. conceived the study.
X.C. performed calculations and numerical simulations.
X.C., C.-R.C., and J.-Q.Z., analyzed the results.
X.C., C.-R.C., J.-Q.Z., and W.-L.Y. wrote and proofread the manuscript.

\subsection*{Declaration of Interests}
The authors declare no competing interests.



\end{document}